\documentstyle[11pt]{article}
\begin{document}
\begin{titlepage}
\begin{flushright}
{OUTP-97-32-P}\\
{July 1997}\\
{to appear in Phys.Lett. B} 
\end{flushright}
\vskip 0.5 cm
\begin{center}
 {\Large{\bf{Strong Unification}}}\\
        \vskip 0.6 cm
        {\large{\bf Dumitru Ghilencea$^{*}$\footnote{
                E-mail address: D.Ghilencea1@physics.oxford.ac.uk}
                ,
                Marco Lanzagorta$^{\dagger}$\footnote{
                E-mail address: unisol@ictp.trieste.it}
                ,
                Graham G. Ross$^{*}$\footnote{
                E-mail: G.Ross1@physics.oxford.ac.uk}}}\\
        \vskip 1 cm
{$^{*}${\it Department of Physics, 
            Theoretical Physics, 
            University of Oxford}}\\
{\it 1 Keble Road, Oxford OX1 3NP, United Kingdom}\\
\vskip 0.2 cm
{$^{\dagger}$ {\it High Energy Section, 
                   International Centre for
                   Theoretical Physics}}\\
             {{\it PO Box 586, Trieste, Italy}}\\
\end{center}
\vskip 2 cm
                                             
\begin{abstract}
We investigate the possibility that unification occurs at strong coupling.
We show that, despite the fact the couplings pass through a strong coupling
regime, accurate predictions for their low energy values are possible because
the couplings of the theory flow to infrared fixed points. We determine the
low-energy QCD coupling in a favoured class of strong coupling models and
find it is reduced from the weak coupling predictions, lying close to the
experimentally measured value. We extend the analysis to the determination
of quark and lepton masses and show that (even without Grand Unification)
the infra-red fixed point structure may lead to good predictions for the top
mass, the bottom to tau mass ratio and $\tan \beta $. Finally we discuss the
implications for the unification scale finding it to be increased from the
MSSM value and closer to the heterotic string prediction.

\end{abstract}
\end{titlepage}

\setcounter{footnote}{0}
\section{Introduction}

The remarkable agreement of the unification predictions for gauge couplings
offers the best evidence for a stage of unification of the fundamental
forces. Further, the determination of the unification scale close to the
(post)diction of string theory may be the first quantitative indication of
unification with gravity. However, in detail, the predictions do not quite
fit with our expectations, particularly in the case of superstring
unification. The evolution of the gauge couplings, with the assumption of
the minimal MSSM spectrum, yields a value for the strong coupling, $%
\alpha_3^o(M_Z)\approx 0.126$, rather high when compared with the
latest average of experimental determinations \cite{exp}, 
$\alpha_3(M_Z)=0.118\pm0.003$. Further, the
unification scale, $M_g^o$, is found to be $(1-3)\times 10^{16}GeV$, 
a factor of 20 below the string scale, which is the typical
expectation for the weakly coupled heterotic string. The unified 
coupling at the unification scale is given by 
$\alpha _{g}^o\approx 0.043$, so the physics around the unification
scale $M_{g}^{o}$ lies in the perturbative domain, but it has been
argued \cite{dine} that this is not acceptable in string theory, as the theory
suffers from the ``dilaton runaway problem''. In order to stabilise the
dilaton one must appeal to non-perturbative effects and the authors of
ref \cite{dine} and \cite{babu} argue that an intermediate value of 
$\alpha_{g}\approx 0.2$ at $M_{g}$ is desired. 
This, they argue, is large enough to stabilize the dilaton, yet
remains perturbative in the sense of quantum field theory \footnote{
String perturbation series are more divergent than field theory series, so
small (perturbative) couplings in QFT can generate large (non-perturbative)
in string theory.} to justify the perturbative analysis of the coupling
unification.

Ways to eliminate these problems have been widely studied\footnote{For
an extensive review see \cite{dienes1}}. Witten 
has found \cite{horava} that the (10 dimensional) strongly coupled
heterotic string theory (M theory) gives a prediction for the unification
scale more closely in agreement with the gauge unification value found in
the MSSM. However, this does not by itself address the problem of dilaton
stability. Stimulated in part by this problem, we have recently explored the
case of unification at intermediate values of gauge coupling at the
unification scale, for which case perturbation theory still may be used up
to the unification scale. We found \cite{grl} that the prediction for the
strong coupling constant and the unification scale is remarkably insensitive
to the addition of massive 
states\footnote{We neglected there the Yukawa effects of the third
generation, but the result is still true when they are included, at least 
for the unification scale \cite{unifpred}.}
which lead to unification at intermediate
coupling. In this letter we extend this analysis in two ways. We discuss in
detail the prospects for obtaining precision predictions for the case the
unified coupling becomes strong. We also consider the implications for
Yukawa couplings which are fixed because they flow rapidly to fixed points
in the case of unification at strong coupling.

One may easily achieve unification at strong coupling through the addition
of a number of additional (massive) multiplets to the MSSM spectrum. It is
notable that such additional multiplets occur in the majority of string
compactifications, coming in representations vectorlike with respect to the
Standard Model gauge group, and hence, likely to acquire mass at the first
stage of spontaneous breaking below the compactification scale, which is
likely to be much higher than the electroweak breaking scale. Thus, it is
reasonable to argue that the MSSM is not the typical case and that we should
consider models with additional massive states as standard. However, this
seems to destroy the success of the unification predictions which are very
sensitive to the addition of such states. In ref \cite{grl} we argued that
this is not the case for the most promising 
extensions of the Standard Model
have additional states which fill out complete $SU(5)$ multiplets and these
do not change the MSSM  unification predictions at one loop order;
this is clearly the case for the case of Grand Unification with a 
Grand Unified group which contains $SU(5)$ for there may easily be 
additional Grand Unified mutiplets with mass below the unification
scale.  However, it may also be the case for
superstring unification, even though the gauge group below the
compactification scale is not Grand Unified. In ref. \cite{witold} it was
argued that level-one string theories with symmetry breaking by Wilson lines
provide the most promising superstring compactified models. In these, the
coupling constant of the various gauge couplings have the usual $SU(5)$
unified values, even though the gauge group below
compactification may naturally need not be Grand Unifed; indeed\footnote{For a
fuller discussion of the possibilities see \cite{dienes2}}, it may be 
just the Standard Model gauge group.
Furthermore, level-one string theories allow only low lying representations
to occur, immediately explaining why quarks and leptons belong only to
triplets of $SU(3)$ and doublets of $SU(2)$. In such models, with
Wilson line breaking, the multiplet structure is 
 predicted to have the generations filling out complete $SU(5)$
representations\footnote{Even though the gauge group is not $SU(5)$.},
just as is observed. Moreover, there is a natural explanation for the 
Higgs doublets of the MSSM, because there are predicted to be at least
two (split) multiplets which do {\it not} fill out complete $SU(5)$ 
representations. There is only a discrete number
of Wilson lines possible and for one of these the split multiplet contains
just the Higgs of the MSSM. The only ambiguity in this class of models is
that it may contain an additional $n_5$ multiplets filling out complete $5+%
\bar{5}$ representations and $n_{10}$ complete 
$10+\overline{10}$ representations.

With this motivation, we now proceed to consider the possibility that gauge
unification occurs at strong coupling. We will show that this leads to
precise unification predictions and will compute them for the class of
models just discussed, in which the additional states leading to strong
coupling unification fill out complete $SU(5)$ representations. We will
further show that such models have a very interesting consequence for
fermion masses because Yukawa couplings may lie in the domain of attraction
of an infra red fixed point of the theory and, due to the strong coupling at
unification, they flow very quickly to the fixed point. This leads to
detailed predictions for the third generation masses.

\section{Strong Unification}

Unification at strong coupling was proposed a long time ago \cite{maiani} as
a viable possibility leading to reasonable predictions for the low-energy
gauge couplings. Here we reformulate the idea in a way that quantifies the
uncertainties in the predictions and refers only to evolution of the
couplings once they reach the {\it perturbative} domain. At first sight it seems
that strong unification does not lend itself to a precise prediction of the
gauge couplings, due to the need to determine the evolution in the strong
coupling domain. The reason this is not the case is because the ratio of
gauge couplings flow to an infrared fixed point. Thus, one has the situation
where the boundary conditions for the evolution of couplings in the MSSM are
still reliably calculable - at the ``intermediate'' mass scale, $M$, of the
new vectorlike states, the ratios of the gauge couplings are given by their
infra-red fixed point values, corresponding to the theory above this mass
scale. For the case the coupling is initially large, the ratio of couplings
closely approaches the fixed point, so a determination of the low energy
values of the couplings, using these boundary conditions plus two-loop MSSM
evolution provides an accurate determination of the couplings.

The two loop renormalisation group equations for the gauge couplings, with
no Yukawa interaction, are given by: 
\begin{equation}  
\label{rge}
\frac{d\alpha_i}{dt}={\tilde b_i}\alpha_i^2+\frac{1}{4\pi} \sum_{j=1}^{3} {%
\tilde b_{ij}}\alpha_i^2\alpha_j+{\cal O}(\alpha^4)
\end{equation}
with $i=\{1,2,3\}$ and where $t=\frac{1}{2\pi}\ln{Q/M_g}$; $M_g$ is the
unification scale, and $\tilde b_i$ and $\tilde b_{ij}$ are the appropriate
one loop and two loop beta functions respectively. To exhibit the
infra-red-fixed-point (IRFP) structure of these equations we rewrite this
equation in the form 
\begin{equation}
\frac{d }{dt}\ln\frac{\alpha_i}{\alpha_k}={\tilde b_i}\alpha_i -{\tilde b_k}%
\alpha_k+ \frac{1}{4\pi}\sum_{j=1}^{3} \left(\tilde b_{ij}\alpha_i-\tilde
b_{kj}\alpha_k\right) \alpha_j+{\cal O}(\alpha^3)  \label{rgefp}
\end{equation}
At one-loop order the evolution of this ratio clearly has an IRFP stable
fixed point\footnote{%
Provided, of course, $\tilde b_i >0$ which is necessary if all the couplings
are to become large at some high scale.} with the fixed point value given by 
\begin{equation}
\left (\frac{\alpha_i}{\alpha_k}\right )^*=\frac{\tilde b_k}{\tilde b_i}
\end{equation}

Provided the gauge couplings are small, the two-loop corrections and above
will only give a small correction to this fixed point value.
Phenomenologically, this must be the case for, to be viable, the couplings
should match at the scale $M$ the values of the low energy couplings evolved
up in energy using the usual MSSM renormalisation group
equations. Provided 
$M$ is large enough, the values of the couplings are all small (of $O(1/24)$
for $M$ near $10^{16}GeV$). In the class of models explored here the two
loop corrections are further suppressed. This follows because above the
intermediate scale the one loop beta functions in the presence of complete $%
SU(5)$ multiplets is given by:

\begin{equation}
\tilde{b}_{i}=b^{\prime}_i=\left( 
\begin{array}{r}
\frac{33}{5}+n \\ 
1+n \\ 
-3+n
\end{array}
\right)  \label{beta}
\end{equation}
where $n$ represents the linear combination
\footnote{here $n_5=N_5+N_{\overline 5}$ and 
$n_{10}=N_{10}+N_{\overline{10}}$}
$n=(n_{5}+3n_{10})/2$. We are
particularly interested in the case $n$ is large for then $M$ is large and
the couplings are driven rapidly to the fixed point ratios. However the
two-loop corrections do not grow with $n$ because, as is clear from \cite
{shifman} (cf eqs. (22),(23), (24)), massive states do not contribute 
two-loop corrections, the usual two-loop contribution to the beta
function being cancelled by the massive threshold corrections
\cite{grl}. Thus the two loop effects are suppressed
both by the additional power of the coupling at the matching scale, $M$, and
by the factor $1/n.$ We shall investigate the magnitude of these corrections
in Section 4.

To summarise we have reformulated the unification of gauge couplings for the
case that unification occurs at large coupling via boundary conditions for
renormalisation group equations which apply {\it below} the scale of new
physics beyond the MSSM. The advantage of this is twofold. It requires
integration of the renormalisation group equations only in the domain where
the coupings are small and perturbation theory applies. It quantifies the
uncertainties in the analysis. The latter come from the two loop and higher
corrections at the matching scale where these are small; hence the
possibility of making accurate predictions for the gauge couplings at low
energies even in the case of unification at strong coupling. Note that
strong unification {\it does not} even require the equality of couplings at any
scale. In this sense the infra-red structure of the theory substitutes for
Grand Unified relations.

\subsection{One loop analysis}

It is instructive to determine the ``strong unification'' predictions at
one-loop order, to show the general trend, before presenting the results of
the full two loop analysis. Below $M$, the multiplet structure is just that
of the MSSM with one-loop beta functions $b_{i}$ given by eq(\ref{beta})
with $n=0$, $\tilde{b_i}=b_{i}=b^{\prime}_{i}(n=0)$. Above $M$ the beta
functions are $b^{\prime}_{i}$. The boundary conditions for the evolution
below the scale $M$ are just 
\begin{equation}
\frac{\alpha_{k}(M)}{\alpha_{i}(M)}=\frac{b^{\prime}_{i}}
{b^{\prime}_{k}}  
\label{ratio}
\end{equation}
From eq(\ref{rge}) we have 
\begin{equation}
\alpha _{i}^{-1}(M_{Z})=\alpha _{i}^{-1}(M)+\frac{b_{i}}{2\pi }\ln \left[ 
\frac{M}{M_{Z}}\right]
\end{equation}
Using this and the boundary condition gives 
\begin{equation}
\frac{1}{2\pi }\ln \left[ \frac{M}{M_{Z}}\right] =\frac{b_{k}^{\prime
}\alpha _{i}^{-1}(M_{Z})-b_{i}^{\prime }\alpha _{k}^{-1}(M_{Z})}{%
(b_{i}-b_{k})n}  \label{t}
\end{equation}
Further, it is straightforward to show that: 
\begin{equation}
(b_{1}-b_{2})\alpha _{3}^{-1}(M_{Z})=(b_{1}-b_{3})\alpha
_{2}^{-1}(M_{Z})+(b_{3}-b_{2})\alpha _{1}^{-1}(M_{Z})  \label{alfas}
\end{equation}
Using this relation one may obtain $\alpha _{3}(M_{Z})$ given the
experimental measurements of the other two couplings. This relation is
identical to that one obtains from the MSSM RGE equations showing that, at
one loop order, the predictions are the same in the MSSM and the
fixed-point-boundary-condition (FPBC) scheme. Note eq(\ref{alfas}) is
independent of the value of $n$ because complete $SU(5)$ multiplets
contribute equally to the one-loop beta functions. As a result, at one loop,
the prediction for $\alpha _{3}(M_{Z})$ in ``strong unification'' is
universal for the class of models with additional complete $SU(5)$
multiplets. With the measured values of $\alpha _{1}(M_{Z})$ and $\alpha
_{2}(M_{Z})$ as input, the value for $\alpha _{3}(M_{Z})$ is $0.1145$. Of
course, two loop and SUSY threshold corrections should be added to obtain a
precision prediction; these will be considered in the next section.

The value of the mass $M$ of the vectorlike states, additional to the MSSM
spectrum, may be obtained from eq(\ref{t}). Taking $i=1$ and $k=2$ one finds 
\begin{equation}
\frac{1}{2\pi }\ln \left[ \frac{M}{M_{Z}}\right]\approx\frac{29.3n-136.9}{5.6n}
\end{equation}
From this, we find that solutions are possible only for $n\ge 5$. The value
of M is clearly $n$ dependent, and varies from $10^{3}GeV$ for $n=5$ to $%
10^{13}GeV$ for $n=20$ and to $10^{16}$ for $n=300$. However, this should
not be interpreted as the normal unification scale at which the couplings
are equal. The latter point occurs in the strong coupling domain, and thus
cannot be precisely determined in perturbation theory. However, one may
determine the scale at which the couplings enter the non-perturbative
domain. At this point they are evolving rapidly, so it is a reasonable
conjecture that they become equal very close to this scale. Remarkably, the
scale $M_{NP}$, at which the couplings become large, turns out to be almost
independent of $n$ and is given by $M_{NP}\approx 3.10^{16}GeV$, essentially
the same scale as is found in the MSSM for the unification scale!

\subsection{Two loop analysis}

In this section we present a two loop analysis in which we compute the scale 
$M$ at which the couplings are in the fixed point ratio as well as the
prediction for $\alpha _{3}(M_{Z})$. The {\it real} unification scale, if 
{\it any}, will not be an output of the scheme.

As usual in making a prediction for the low energy values of the gauge
couplings we are faced with the problem of unknown values for the
supersymmetric spectrum; this can significantly affect the predictions we
make because, at two loop order, one has to take into account the low energy
supersymmetric thresholds in one loop. Various scenarios for low
supersymmetric energy spectrum have been studied in the minimal
supersymmetric standard model (MSSM), and their effect on the value of the
unification scale as well as $\alpha _{3}^{o}(M_{Z})$ has been extensively
discussed \cite{langaker}. Given this we choose to make our predictions {\it %
relative} to the MSSM\ prediction calculated with a given SUSY threshold. 
In the MSSM we have

\begin{equation}
\alpha _{i}^{o-1}(M_{z})=-\delta _{i}^{o}+\alpha _{g}^{o-1}+\frac{b_{i}}{%
2\pi }\ln \left[ \frac{M_{g}^{o}}{M_{Z}}\right] +\frac{1}{4\pi }%
\sum_{j=1}^{3}\frac{b_{ij}}{b_{j}}\ln \left[ \frac{\alpha _{g}^{o}}{\alpha
_{j}^{o}(M_{Z})}\right]  \label{MSSM}
\end{equation}
where the $\delta _{i}^{o}$ contains the effect of the low energy
supersymmetric thresholds and we have ignored the Yukawa couplings effects,
which are known to be small. The MSSM variables are labelled with an ``o''
index to distinguish them from the model based on fixed point scenario.

In the FPBC case we have

\begin{equation}
\alpha _{i}^{-1}(M_{z})=-\delta _{i}^{o}+\alpha _{i}^{-1}(M)+\frac{b_{i}}{%
2\pi }\ln \left[ \frac{M}{M_{Z}}\right] +\frac{1}{4\pi }\sum_{j=1}^{3}\frac{%
b_{ij}}{b_{j}}\ln \left[ \frac{\alpha _{j}(M)}{\alpha _{j}(M_{Z})}\right]
\label{FPSM}
\end{equation}
where, as discussed above, we use the same threshold $\delta _{i}^{o}$ as in
the MSSM. The $b_{i}$ and $b_{ij}$ denote the one loop and two loop beta
functions which are just the same as in the MSSM; $M$ denotes the scale
where the couplings $\alpha _{i}(M)$ are in the ``fixed-point'' ratio. Now,
from experiment we have well measured values for $\alpha _{1}(M_{Z})$ and $%
\alpha _{2}(M_{Z})$ from the values of electromagnetic coupling and Weinberg
angle at $M_{Z}$ scale. Therefore, in computing $\alpha _{3}(M_{Z})$, these
values were taken as input from experiment. The FPBC case should comply with
this condition, too, and hence $\alpha _{1}(M_{Z})=\alpha _{1}^{o}(M_{Z})$
and $\alpha _{2}(M_{Z})=\alpha _{2}^{o}(M_{Z})$.

We subtract the eqs. (\ref{MSSM}), (\ref{FPSM}) and, using these relations,
obtain

\begin{equation}
0=\alpha _{i}^{-1}(M)-\alpha _{g}^{o-1}+\frac{b_{i}}{2\pi }\ln \left[ \frac{M%
}{M_{g}^{o}}\right] +\frac{1}{4\pi }\sum_{j=1}^{3}\frac{b_{ij}}{b_{j}}\ln
\left[ \frac{\alpha _{j}(M)}{\alpha _{g}^{o}}\right] +\frac{1}{4\pi }\frac{%
b_{i3}}{b_{3}}\ln \left[ \frac{\alpha _{3}^{o}(M_{Z})}{\alpha _{3}(M_{Z})}%
\right]  \label{res1}
\end{equation}
for the case $i=\{1,2\}$ and 
\begin{equation}
\alpha _{3}^{-1}(M_{Z})-\alpha _{3}^{o-1}(M_{Z})=\alpha _{3}^{-1}(M)-\alpha
_{g}^{o-1}+\frac{b_{3}}{2\pi }\ln \left[ \frac{M}{M_{g}^{o}}\right] +\frac{1%
}{4\pi }\sum_{j=1}^{3}\frac{b_{3j}}{b_{j}}\ln \left[ \frac{\alpha _{j}(M)}{%
\alpha _{g}^{o}}\right] +\frac{1}{4\pi }\frac{b_{33}}{b_{3}}\ln \left[ \frac{%
\alpha _{3}^{o}(M_{Z})}{\alpha _{3}(M_{Z})}\right]  \label{res2}
\end{equation}
for the case $i=3.$ At two loop order we may make the approximation 
\[
\ln \left[ \frac{\alpha _{3}^{o}(M_{Z})}{\alpha _{3}(M_{Z})}\right] =\ln
\alpha _{3}^{-1}(M_{Z})_{oneloop}-\ln \alpha _{3}^{o-1}(M_{Z})_{oneloop}=0 
\]
One may readily check that the LHS is indeed numerically very small. This
gives 
\begin{equation}
0=\alpha _{i}^{-1}(M)-\alpha _{g}^{o-1}+\frac{b_{i}}{2\pi }\ln \left[ \frac{M%
}{M_{g}^{o}}\right] +\frac{1}{4\pi }\sum_{j=1}^{3}\frac{b_{ij}}{b_{j}}\ln
\left[ \frac{\alpha _{j}(M)}{\alpha _{g}^{o}}\right]
\end{equation}
for the case $i=\{1,2\}$ and 
\begin{equation}
\alpha _{3}^{-1}(M_{Z})-\alpha _{3}^{o-1}(M_{Z})=\alpha _{3}^{-1}(M)-\alpha
_{g}^{o-1}+\frac{b_{3}}{2\pi }\ln \left[ \frac{M}{M_{g}^{o}}\right] +\frac{1%
}{4\pi }\sum_{j=1}^{3}\frac{b_{3j}}{b_{j}}\ln \left[ \frac{\alpha _{j}(M)}{%
\alpha _{g}^{o}}\right]
\end{equation}
for $i=3.$

This is a system of three equations with the unknowns: $\alpha _{3}(M_{Z})$, 
$\,M$ and one of the $\alpha _{i}(M)$'s (say $\alpha _{1}(M)$), as the ratio
of any two of them is a known function for any given $n$ through eq(\ref
{ratio}). The solution is

\begin{equation}
\alpha _{3}^{-1}(M_{Z})=\alpha _{3}^{o-1}(M_{Z})-\frac{470}{77\pi }\ln
\left[ \frac{\alpha _{1}(M)}{\alpha _{g}^{o}}\right] +\frac{17}{14\pi }\ln
\left[ \frac{b_{1}+n}{b_{3}+n}\right] -\frac{15}{2\pi }\ln \left[ \frac{%
b_{1}+n}{b_{2}+n}\right]
\end{equation}
for the strong coupling and 
\begin{equation}
\ln \left[ \frac{M}{M_{g}^{o}}\right] =-\frac{2\pi }{n\alpha _{g}^{o}}+{%
\frac{(2336+341n)}{231n}}\ln \left[ \frac{\alpha _{1}(M)}{\alpha _{g}^{o}}%
\right] +{\frac{57+7n}{4n}}\ln \left[ \frac{b_{1}+n}{b_{2}+n}\right]
-\frac{4(22+n)}{21n}\ln\left[\frac{b_1+n}{b_3+n}\right]
\end{equation}
for the scale M, where the value of $\alpha _{1}(M)$ is given by the root of
the nonlinear equation: 
\begin{equation}
\alpha _{1}^{-1}(M)=\frac{b_{1}+n}{n\alpha_{g}^{o}}-\frac{1168}{231\pi }%
\frac{(b_{1}+n)}{n}\ln \left[ \frac{\alpha _{1}(M)}{\alpha _{g}^{o}}\right] -%
\frac{57}{8\pi }\frac{(b_{1}+n)}{n}\ln \left[ \frac{b_{1}+n}{b_{2}+n}%
\right]
+\frac{44}{21 \pi}\frac{b_1+n}{n}\ln\left[\frac{b_1+n}{b_3+n}\right]  
\label{alfa}
\end{equation}
with $b_{1}=33/5$,$\,b_{2}=1$,$\,b_{3}=-3$.

Using these expressions we get the numerical results presented in Table 1 in
which we have taken as the reference MSSM prediction $\alpha
_{3}^{o}(M_{Z})=0.126$. We see we get a lower value for alpha strong
at electroweak scale than in the MSSM and closer to the experimental
measurement \cite{exp} $\alpha _{3}(M_{Z})=0.118\pm 0.003$. Since the
couplings are quite small at the high scale $M$, the higher corrections to
the boundary conditions are expected to be small. We will estimate these
corrections in Section 4. 
\begin{table}[tbp]
\begin{center}
\begin{tabular}{|c|c|c|c|c|c|c|}
\hline
$n$ & $\alpha_1(M)$ & $\alpha_3(M_Z)$ & $M$ & $\frac{m_b}{m_{\tau}}(M_Z) $ & 
$m_t$ & $\tan\beta$ \\ \hline\hline
6 & 0.020 & 0.1163 & $2.14\times 10^5 $ & 1.62 & 229.26 & 47.15 \\ \hline
8 & 0.023 & 0.1188 & $1.61\times 10^8 $ & 1.62 & 209.87 & 46.99 \\ \hline
10 & 0.025 & 0.1203 & $0.79\times 10^{10}$ & 1.61 & 204.06 & 46.82 \\ \hline
12 & 0.027 & 0.1213 & $1.04\times 10^{11}$ & 1.60 & 202.58 & 46.70 \\ \hline
14 & 0.029 & 0.1219 & $6.44\times 10^{11}$ & 1.59 & 201.51 & 46.64 \\ \hline
16 & 0.030 & 0.1224 & $2.52\times 10^{12}$ & 1.58 & 200.92 & 46.54 \\ \hline
18 & 0.031 & 0.1228 & $7.23\times 10^{12}$ & 1.57 & 200.57 & 46.48 \\ \hline
20 & 0.032 & 0.1231 & $1.68\times 10^{13}$ & 1.57 & 200.34 & 46.43 \\ \hline
22 & 0.033 & 0.1234 & $3.34\times 10^{13}$ & 1.56 & 200.20 & 46.41 \\ \hline
26 & 0.034 & 0.1238 & $0.96\times 10^{14}$ & 1.55 & 200.03 & 46.32 \\ \hline
\end{tabular}
\end{center}
\caption{The value of $\alpha_1$ at the intermediate scale, the strong
coupling at $M_Z$, and the intermediate scale obtained using the fixed-point
boundary conditions as a function of n. Also shown are the bottom to tau
mass ratio and the top mass for the case the third generation couplings are
in the domain of attraction of the fixed point.}
\label{table:1}
\end{table}

\section{Strong Unification and the masses of the third generation}

As we have discussed, the addition of massive multiplets to the MSSM is to
be expected in viable Grand Unified theories and in many string theories.
The effect of such new states is to increase the gauge coupling at
unfication and can easily make it approach the strong coupling domain. We
further remarked that in this domain the fixed point structure of the theory
relating the largest Yukawa couplings (and hence third generation masses) to
the gauge couplings becomes the dominant effect as the couplings flow
rapidly towards the fixed points. In this section we explore these
implications in detail.

The renormalisation group equations for the Yukawa couplings in the MSSM are
given by 
\begin{eqnarray}
\frac{d}{dt}Y_\tau &=& Y_\tau \left(3Y_b+4 Y_\tau-\frac{9}{5} \alpha_1 
-3\alpha_2\right)\nonumber \\
\frac{d}{dt}Y_b &=& Y_b \left(Y_t+6 Y_b+Y_\tau-\frac{7}{15}\alpha_1-3
\alpha_2-\frac{16}{3}\alpha_3\right)\nonumber\\
\frac{d}{dt}Y_t &=& Y_t \left(6 Y_t+Y_b
-\frac{13}{15}\alpha_1-3\alpha_2-\frac{16}{3}\alpha_3\right)
\label{rgmssm}
\end{eqnarray}
where $Y_j = h_j^2/4\pi$ and $h_j$ is the Yukawa coupling. If we ignore
the smaller gauge couplings $\alpha_1$ and $\alpha_2$ and we keep only the
large top Yukawa coupling $Y_t$, the last equation has the form 
\begin{eqnarray}
\frac{d Y_t}{dt}&=&Y_t(sY_t-r_3 {\alpha_3})  
\label{eq:rg}
\end{eqnarray}

This has a fixed point that relates the top Yukawa coupling to the QCD
coupling $\alpha_3$ given by \cite{PR}, \cite{irsfp}: 
\begin{equation}
\left( \frac{Y_t}{\alpha_3} \right)^* = \frac{r_3 + b_3}{s}
\end{equation}
which is infra-red stable if $r_3+b_3 >0$. In the case of the MSSM, 
$b_3=-3$, $r_i = (13/15,3,16/3)$ and $s=6$, then, the fixed point 
is $\left(\frac{Y_t} {\alpha_3}\right)^*_{MSSM}=\frac{7}{18}$.

However, as stressed in ref. \cite{hill}, this fixed point value is
not reached for large initial values of the top quark coupling because the
range in $t$ between the Planck scale and the electroweak scale is too small
to cause the trajectories to closely approach the fixed point. As
demonstrated in \cite{irsfp} a ``Quasi-fixed point'' governs the value of
$Y_{t} $ for large initial values of $Y_{t}$ and is given by 
\begin{equation}
\left( \frac{Y_{t}}{{\alpha}_{3}}\right) ^{QFP}=\frac{\left( \frac{Y_{t}}{{%
\alpha}_{3}}\right) ^{*}}{\left( 1-\left( \frac{{\alpha}_{3}(t)}{{\alpha}%
_{3}(0)}\right) ^{B_{3}}\right) }
\end{equation}
where $B_{3}=\frac{r_{3}}{b _{3}}+1$. The term 
$\left( \frac{{\alpha}_{3}(t)}{{\alpha}_{3}(0)}\right) ^{B_{3}}$ 
determines the rate of approach to the fixed point, the smaller this 
term, the closer the QFP is to the IRSFP. For
the MSSM in the case only the top Yukawa coupling is in the domain of
attraction of the fixed point, including electroweak corrections, the
quasi-fixed point predicts a top quark mass of $m_{t}\approx 210\sin \beta $ 
$GeV$, where $\tan \beta $ is the ratio of the vacuum expectation values of
the Higgs doublets. For the case that the top and the bottom Yukawa
couplings are in the domain of attraction of the fixed point, the prediction
for the fixed point is $m_{t}\approx 190GeV,$ and the dependence on $\beta $
disappears in this case, because the top and bottom Yukawas are nearly equal
and thus we are in the region with $\sin $ $\beta \approx 1.$

As has been shown in \cite{irsfp}, the profusion of new fields increase the
rate of approach to the fixed point. This follows because the gauge
couplings are evolving rapidly so the convergence factor $\left( \frac{{%
\alpha}_{3}(t)}{{\alpha}_{3}(0)}\right) ^{B_{3}}$ is very small. Thus  we
can expect that the IRSP structure will play an important role for the
determination of the couplings in the class of models considered in this
paper. In this case one must keep all three gauge couplings, as all are
comparable above the scale $M$. However the analysis is tractable because
the ratios of the gauge couplings are given by the infra-red fixed point
ratio of eq(\ref{ratio}). The renormalisation group equations may be written
in the form (see also ref.\cite{bando}) 
\begin{eqnarray}
\frac{d}{dt}\ln\left( \frac{Y_{\tau }}{\alpha _{i}}\right) &=&\alpha
_{i}\left( 4\frac{Y_{\tau }}{\alpha _{i}}+3\frac{Y_{b}}{\alpha _{i}}-\frac{9%
}{5}\frac{\alpha _{1}}{\alpha _{i}}-3\frac{\alpha _{2}}{\alpha _{i}}%
-(b_{i}+n)\right)  \label{rgep} \\
&&  \nonumber \\
\frac{d}{dt}\ln\left( \frac{Y_{b}}{\alpha _{i}}\right) &=&\alpha _{i}\left( 
\frac{Y_{\tau }}{\alpha _{i}}+6\frac{Y_{b}}{\alpha _{i}}+\frac{Y_{t}}{\alpha
_{i}}-\frac{7}{15}\frac{\alpha _{1}}{\alpha _{i}}-3\frac{\alpha _{2}}{\alpha
_{i}}-\frac{16}{3}\frac{\alpha _{3}}{\alpha _{i}}-(b_{i}+n)\right) \\
&&  \nonumber \\
\frac{d}{dt}\ln\left( \frac{Y_{t}}{\alpha _{i}}\right) &=&\alpha _{i}\left( 
\frac{Y_{b}}{\alpha _{i}}+6\frac{Y_{t}}{\alpha _{i}}-\frac{13}{15}\frac{%
\alpha _{1}}{\alpha _{i}}-3\frac{\alpha _{2}}{\alpha _{i}}-\frac{16}{3}\frac{%
\alpha _{3}}{\alpha _{i}}-(b_{i}+n)\right)
\end{eqnarray}
where the index $i$ is fixed. Using eq.(\ref{ratio}) we get the following
fixed points: 
\begin{eqnarray}
\left(\frac{Y_{\tau }}{\alpha _{i}}\right) ^{*} &=&b_{i}^{\prime }\left( 
\frac{10}{61}+\frac{143}{305}\frac{1}{b_{1}^{\prime }}+\frac{30}{61}\frac{1}{%
b_{2}^{\prime }}-\frac{40}{61}\frac{1}{b_{3}^{\prime }}\right) \\
&&  \nonumber \\
\left(\frac{Y_{t}}{\alpha _{i}}\right)^{*} &=&b_{i}^{\prime }\left(
\frac{9}{61}+\frac{136}{915}\frac{1}{b_{1}^{\prime }}+\frac{27}{61}\frac{1}{%
b_{2}^{\prime }}+\frac{136}{183}\frac{1}{b_{3}^{\prime }}\right) \\
&&  \nonumber \\
\left(\frac{Y_{b}}{\alpha_{i}}\right)^{*} &=&b_{i}^{\prime }\left( 
\frac{7}{61}-\frac{23}{915}\frac{1}{b_{1}^{\prime }}+\frac{21}{61}\frac{1}{%
b_{2}^{\prime }}+\frac{160}{183}\frac{1}{b_{3}^{\prime }}\right)
\end{eqnarray}
These fixed point ratios apply at the scale $M$. Below this scale the
couplings evolve via the usual MSSM renormalisation group equations, eq(\ref
{rgmssm}). Using these fixed point boundary conditions we have integrated
these equations numerically, including a SUSY\ threshold at $300GeV.$ This
allows us to determine $m_{t}$, the ratio $m_{b}/m_{\tau }$ 
and $\tan \beta$ at the $M_{Z}$ scale for the large $\tan\,\beta$ 
case. The results are presented in Table 1, with an acceptable value 
\cite{deboer} for  $m_{b}/m_{\tau }$ and a rather high value for $m_t$.
For the low $tan \beta$ case 
\footnote{This case arises if the light doublet giving mass to 
the down quarks and leptons is a mixture of two (or more) of the 
Higgs fields, the dominant component not coupling to the quarks and 
leptons. Such mixing is discussed in more detail below.}
we determine $m_{t} \sin\beta $, which is
approximately 1.08 times the value
for $m_{t}$ given in Table 1. We see that the prediction 
for $m_{t} \sin \beta$ is very
high. Thus we are driven to the low $\tan $ $\beta $ solution with $\tan $ $%
\beta $ in the range (1.01-1.3). 


\begin{table}[tbp]
\begin{center}
\begin{tabular}{|c||c|c|}
\hline
$n$ & $\frac{m_b}{m_{\tau}}(M_Z) $ & $m_t$ \\ \hline\hline
6 & 1.41 & 184.58 \\ \hline
8 & 1.53 & 185.85 \\ \hline
10 & 1.57 & 187.69 \\ \hline
12 & 1.59 & 189.20 \\ \hline
14 & 1.59 & 190.42 \\ \hline
16 & 1.59 & 191.38 \\ \hline
18 & 1.59 & 192.17 \\ \hline
20 & 1.59 & 192.83 \\ \hline
22 & 1.59 & 193.40 \\ \hline
26 & 1.58 & 194.27 \\ \hline
\end{tabular}
\end{center}
\caption{The bottom to tau mass ratio and top mass as a function of $n$ for
the case all three generations have equivalent coupling to Higgs states.}
\label{table:2}
\end{table}

\begin{table}[tbp]
\begin{center}
\begin{tabular}{|c||c|c|}
\hline
$n$ & $\frac{m_b}{m_{\tau}}(M_Z) $ & $m_t$ \\ \hline\hline
6 & 1.43 & 128.28 \\ \hline
8 & 1.59 & 144.44 \\ \hline
10 & 1.65 & 154.19 \\ \hline
12 & 1.67 & 160.82 \\ \hline
14 & 1.68 & 165.68 \\ \hline
16 & 1.67 & 169.40 \\ \hline
18 & 1.67 & 172.36 \\ \hline
20 & 1.67 & 174.78 \\ \hline
22 & 1.66 & 176.80 \\ \hline
26 & 1.65 & 179.98 \\ \hline
\end{tabular}
\end{center}
\caption{The bottom to tau mass ratio and top mass as a function of $n$ for
the couplings of Table 2 plus additional couplings involving heavy doublet
quarks and leptons.}
\end{table}

The values for the third generation masses are rather sensitive to the
number of Yukawa couplings lying in the domain of attraction of the
infra-red fixed points. The case just presented corresponds to a
near-minimal case where only the third generation Yukawas are large. However,
with the addition of numerous vectorlike multiplets, there are many more
possible Yukawa couplings which can affect our conclusions. To illustrate
this we first consider a model in which each of the quark and lepton
families has a coupling to a (heavy) Higgs state, corresponding to the terms
in the superpotential $\sum_{i,j=1}^{3}($ $h_{ij}^{u}Q_{i}U_{j}H_{2}^{ij}$+$%
h_{ij}^{d}Q_{i}U_{j}H_{1}^{ij}+h_{ij}^{l}L_{i}e_{j}H_{1}^{ij}).$ Such a
model has been proposed as a means of dynamically solving the SUSY flavour
problem \cite{LR} and for generating structure in the light quark mass
matrix \cite{ross} via mixing in the Higgs sector so that the two light
Higgs doublets of the MSSM are mixtures of $H_{1}^{ij}\,$and of $H_{2}^{ij}.$
In this case the renormalisation group equations are straightforward
generalisations of eq(\ref{rgep}). Solving them for the fixed points and
using these as boundary conditions for the MSSM renormalisation group
equations gives the results shown in Table 2. Finally, we consider a model in
which there are also Yukawa couplings involving mixing of the quarks and
leptons with the new quark and lepton states belonging to the vectorlike
representations. We restrict our example to the case of couplings between
additional $u_{L}$, $b_{L}$ and $\tau _{L}$ states to the Higgs and
right-handed quarks and leptons via the terms 
$\sum_{i,j=1}^{3}(h_{ij}^{u\prime }Q_{i}^{\prime }U_{j}H_{2}^{ij}$
+$h_{ij}^{d\prime}Q_{i}^{\prime }U_{j}H_{1}^{ij}
+h_{ij}^{l\prime }L_{i}^{\prime}e_{j}H_{1}^{ij})$. 
The results for this case are presented in Table 3. We
see from these tables that the result for the ratio $m_{b}/m_{\tau }\,$is
quite stable and remains acceptable for most values of $n$. The result
for $m_{t}$ is quite sensitive to the number of Yukawa couplings and to $n$. We see
that even for the large $\tan\beta \,$case when the top mass has a
definite prediction such models can lead to a remarkably consistent pattern
of third generation masses.

\section{Corrections to FPBC formalism}

To discuss the corrections to the fixed-point boundary condition formalism,
it is most convenient to work with the form Shifman \cite{shifman}
derived for the running couplings 
\begin{eqnarray}
\alpha _{1}^{-1}(M) &=&\alpha _{g}^{-1}+\frac{1}{2\pi }%
\frac{3}{10}\left\{ \sum_{gen}{}\left[ \ln \frac{M_{g}}{MZ_{lL}}+2\ln 
\frac{M_{g}}{MZ_{eR}}+\frac{1}{3}\ln \frac{M_{g}}{MZ_{qL}}\right.
\right.  \nonumber \\
&&  \nonumber \\
&&+\left. \left. \frac{8}{3}\ln \frac{M_{g}}{MZ_{uR}}+\frac{2}{3}\ln 
\frac{M_{g}}{MZ_{dR}}\right] +2\ln \frac{M_{g}}{MZ_{H_{u,d}}}\right\} +%
\frac{n}{2\pi }\ln \frac{M_{g}}{\mu _{g}}
\end{eqnarray}

\begin{eqnarray}
\alpha_2^{-1}(M)&=&\alpha_g^{-1}-\frac{6}{2\pi}\ln\frac{M_g}{M
\left(\frac{\alpha_g}{\alpha_2(M)}\right)^{1/3}} +\frac{1}{2\pi}%
\sum_{gen}{}\left[\frac{3}{2}\ln\frac{M_g}{M Z_{qL}} +\frac{1}{2}\ln\frac{%
M_g}{M Z_{lL}}\right]  \nonumber\\
\nonumber \\
& & +\frac{1}{2\pi}\ln\frac{M_g}{M Z_{H_{u,d}}} +\frac{n}{2\pi}\ln\frac{M_g}{\mu_g}
\end{eqnarray}

\begin{eqnarray}
\alpha _{3}^{-1}(M) &=&\alpha _{g}^{-1}-\frac{9}{2\pi }\ln 
\frac{M_{g}}{M\left( \frac{\alpha _{g}}{\alpha _{3}(M)}\right) ^{1/3}%
}+\frac{1}{2\pi }\sum_{gen}{}\left[ \ln \frac{M_{g}}{MZ_{qL}}+\frac{1}{2}%
\ln \frac{M_{g}}{MZ_{uR}}\right.\nonumber\\
&&  \nonumber \\
&&+\left. \frac{1}{2}\ln \frac{M_{g}}{MZ_{dR}}\right] +\frac{n}{2\pi }%
\ln \frac{M_{g}}{\mu _{g}}
\label{eqShifman}
\end{eqnarray}
The advantage of this form for the running of gauge couplings is that (above
the supersymmetric scale) it is {\it exact} to all
orders. However, the wave function renormalisation coefficients $Z_i$ are
only known perturbatively so one is still confined to a given order in
perturbation theory when testing coupling unification. To two loop order,
one must include the values of wave-function renormalisation coefficients $Z_i$
in one loop only. In this formula the $Z_{i}$ factors are
 evaluated at the scale M.

The one loop form of these equations gives the FPBC formalism used
above. The two loop corrections came from the terms involving 
$\ln(\alpha_g/\alpha_2(M))$ and $\ln(\alpha_g/\alpha_3(M))$, the gauge
wave function renormalisation and also from the terms involving the
$Z_i$'s, the matter wave function renormalisation.
To good approximation the former do not affect the FPBC predictions
for $\alpha_3(M_Z)$. To see this first set  $Z_i=1$.
Now the result follows because the predictions for $\alpha_3(M_Z)$
involves the differences $[\alpha_i^{-1}(M)-\alpha_j^{-1}(M)]$, 
$i,j=\{1,2,3\}$.
In these, the $\ln(M_g/M)$ terms coming from the three
generations cancel, leaving just the gauge wave function
renormalisation terms proportional to
$\ln\left[M_g/(M(\alpha_g/\alpha_i(M))^{1/3})\right]$ and the
Higgs contribution proportional to $\ln(M_g/M)$.
The latter has a small coefficient, so we can effectively absorb
the $\left(\alpha_g/\alpha_i(M)\right)^{1/3}$ term in a redefinition
of $M_g$, $M_g\rightarrow M_g/(\alpha_g/\alpha_i(M))^{1/3}$   
with $\alpha_i(M)\approx$ constant.

Hence, to a good approximation, we obtain the fixed-point boundary 
condition formalism  provided we  interpret 
$M_{g}/\left({\alpha_{g}}/{\alpha_{g}^{0}}\right)^{1/3}$ 
as an {\it effective} scale $M_{g}^{\prime}$. We established this for the
case we set  $Z_i(M)=1$; thus, one expects corrections to FPBC 
formalism if this condition is not respected.
How large are they? There are two contributions to each $Z_i$, one
coming from gauge interactions and one from Yukawa interactions. For
couplings remaining in the perturbative domain, the effects of the gauge
coupling terms alone was analysed in \cite{grl} where it was found that the
value of $\alpha _{3}$ {\it increased} slightly. However, as we have
stressed, in strong coupling one expects the Yukawa couplings to be large
and their contribution to  $Z_i$ factors is opposite to that of gauge
interactions, taking the result closer to the FPBC result. 
We may check whether this happens in the case unification occurs at 
intermediate coupling where the perturbative approach still
applies. In this case we may use the one-loop form for the $Z_i$ factors.
Following the argument presented, we have checked in specific cases
that, because of these cancellations, the prediction does lie close to
the FPBC predicted value.

Finally it is of interest to consider the expectation for the unification
scale. As we have stressed this is not determined in the case the coupling
is really strong at unification. However, for intermediate values, say
of $O(0.3)$, one may use eq(\ref{eqShifman}) to determine $M_{g}$. Again  
we find $M_{g}$ larger than the MSSM value. Part of this increase 
follows simply from the fact noted above that part of the two
loop corrections may be absorbed in a change in the unification scale, 
giving $M_{g}=M_{g}^{\prime}\left(\frac{\alpha_{g}}{\alpha_{g}^{0}}\right)^{1/3}$
(For $\alpha_{g}=1$ this approximately gives an increase by
a factor 3). The remaining effect comes from the two loop corrections
arising from the $Z$ factors. It is interesting that together they take the
value of $M_{g}$ closer to the weakly coupled heterotic string prediction.

\section{Conclusions}

Unification at strong coupling is quite likely in extensions of the MSSM
which contain additional states with mass below the unification scale. Since
such cases are perhaps the norm, it is important to determine their
implications for gauge coupling unification. We have shown that it is
possible to determine these implications with surprising accuracy given the
fact that the gauge coupling evolution involves a stage of strong coupling.
The formulation of the initial boundary conditions at the intermediate scale
in terms of the infra-red fixed point ratios of gauge couplings of the
theory above this scale shows that the uncertainties in gauge coupling
predictions are of two loop order. These corrections are small since they
should be evaluated at the intermediate scale where the couplings are small.
The two loop corrections are further suppressed at large $n$, where $n$
specifies the number of additional states, and also by cancellation between
two loop effects involving gauge and Yukawa couplings. The net result is
that the predicted value of the strong coupling is reduced from the MSSM
value coming closer to the experimental value. For unification at
intermediate coupling in which perturbation theory may be used above the
intermediate scale the unification scale is also raised relative to the MSSM
prediction, taking it closer to the heterotic string prediction. The case of
strong unification also leads to predictions for quark and lepton masses
because the Yukawa couplings are driven towards infra-red fixed points. We
have investigated these predictions and found that they may lead to
excellent predictions for the third generation masses. It may be hoped that
such structure will ultimately shed light on the pattern of light quark and
lepton masses too.

\section{Acknowledgements}
The authors want to  thank G. Amelino-Camelia, D.R.T. Jones, 
I. Kogan, J. March-Russell, M. Shifman for useful discussions.
D.G. gratefully acknowledges the financial support from the part of
University of Oxford and Oriel College (University of Oxford).


\begin{thebibliography}{99}
\bibitem{exp}  Review of Particle Data, Phys. Rev D54 (1996), 83.
\bibitem{dine}  M.Dine and N. Seiberg, Phys. Lett. B 162 (1985), 299;\newline
T. Banks and M. Dine, Phys. Rev D 50, (1994) 7454;\newline
M. Dine and Y. Shirman, Phys. Lett. B377 (1996), 36;\newline
M. Dine, preprint hep-th/9508085.
\bibitem{babu}  K. S. Babu and J.C. Pati, Phys. Lett. B 384 (1996), 140.
\bibitem{dienes1} K.R. Dienes, Phys.Rept. 287 (1997) 447.
\bibitem{horava} 
E. Witten, Nucl. Phys. B471 (1996) 135;\newline
P.Horava, Phys. Rev. D 54 (1996), 7561;\newline
P.Horava and E. Witten, Nucl. Phys. B475 (1996), 94.
\bibitem{grl}  D. Ghilencea, M. Lanzagorta-Saldana, G.G. Ross,
Oxford University preprint, OUTP-97-31-P, hep-ph/9707401.\newline
Previous related works on strong coupling:\newline
C. Kolda and J. March-Russell, hep-ph/9609480;\newline
K. S. Babu and J.C. Pati, Phys. Lett. B 384 (1996), 140;\newline
R. Hempfling, Phys. Lett. B351 (1996) 206;\newline
B. Brahmachari, U. Sarkar and K. Sridhar, Mod. Phys. Lett. A8 (1993), 3349.
\bibitem{unifpred} D. Ghilencea and G.G. Ross, work in preparation.
\bibitem{witold}  W. Pokorski and G.G. Ross, Oxford University preprint,
OUTP-97-34-P, hep-ph/9707402.
\bibitem{dienes2} K.R. Dienes, A.E. Farragi and J. March-Russell,
Nucl. Phys. B467 (1996) 44.
\bibitem{maiani} 
L. Maiani, G. Parisi and R. Petronzio, Nucl. Phys. B136 (1978), 115.\newline
L. Maiani and R. Petronzio, Phys. Lett. B176 (1986), 120.
\bibitem{shifman} M. Shifman, Int. J. Mod. Phys. A11 (1996) 5761 and
references therein.
\bibitem{langaker}  P.Langacker and N. Polonsky, Phys. Rev. D47 (1993), 4028
and references therein.
\bibitem{PR} B. Pendelton and G.G. Ross, Phys. Lett. B98 (1981), 291.
\bibitem{irsfp} M. Lanzagorta and G.G. Ross, Phys. Lett. B349 (1995), 319. 
\bibitem{hill}C.T. Hill, Phys. Rev. D24 (1981) 691; \\
 C.T. Hill, C.N. Leung and S. Rao, Nucl. Phys. B262 (1985) 517.
\bibitem{bando}  M. Bando, J. Sato and K. Yoshioka, Kyoto University
preprint, KUNS-1437 HE(TH) 97/04,  hep-ph/9703321; \\
 M. Bando, T. Onogi, J. Sato and T. Takeuchi, CERN preprint
CERN-TH/96-363, hep-ph/9612493
\bibitem{deboer} W. de Boer et al., hep-ph/9603350
\bibitem{LR} M. Lanzagorta and G. G. Ross, Phys. Lett. B364 (1995), 163.
\bibitem{ross} 
Luis Ibanez and Graham G. Ross, Phys. Lett. B332 (1994) 100;\newline
Graham G. Ross,  Phys. Lett. B364 (1995) 216.
\end{thebibliography}
\end{document}